**Operation Regimes and Design Principles of Delta-E Effect Sensors**

*Fatih Ilgaz, Elizaveta Spetzler\*, Patrick Wiegand, Robert Rieger, Jeffrey McCord, Benjamin Spetzler\**

F. Ilgaz, E. Spetzler, J. McCord, B. Spetzler
Department of Materials Science
Kiel University
Kaiserstr. 2, 24143 Kiel, Germany
E-mail: besp@tf.uni-kiel.de, elgo@tf.uni-kiel.de

P. Wiegand, R. Rieger
Department of Electrical and Information Engineering
Kiel University
Kaiserstr. 2, 24143 Kiel, Germany

J. McCord
Kiel Nano, Surface and Interface Science (KiNSIS)
Kiel University
Christian-Albrechts-Platz 4, 24118 Kiel, Germany

**Abstract**

Delta-E effect-based magnetoelectric sensors have emerged as promising technology for detecting weak magnetic fields at low frequencies. However, the performance of such sensors remains difficult to predict, as signal and noise characteristics are dictated by interdependent parameters such as magnetic layer geometry, magnetic microstructure, and loss. In this work, we present a systematic experimental study of sub-mm-sized delta-E effect sensors, comprising 24 device configurations that vary in magnetic layer thickness and lateral dimensions. The sensors are statistically analyzed to identify the influence of magnetic layer geometry on performance through a combination of measurements and simulations. Our findings reveal three distinct operation regimes - dominated by electronic noise, magnetic noise, and nonlinearities - whose boundaries shift systematically with magnetic layer thickness. This regime behavior governs the trade-offs between sensitivity and noise, ultimately determining the sensor's limit of detection. Based on these results, the dependency of the regime boundaries on key device parameters is discussed in detail, providing fundamental insights for tailoring sensor performance. As such, this study establishes a necessary foundation for targeted performance optimization and the scalable design of advanced delta-E effect sensor systems.

## 1. Introduction

Magnetic field sensors have become important components in a wide range of applications, such as the aerospace industry,[1] automotive systems,[2] consumer electronics,[3,4] navigation technologies,[5] and biomedical applications.[6–9] These sensors operate on various concepts, including magnetoresistance (XMR),[10–12] the Hall effect,[13,14] giant magnetoimpedance (GMI),[15,16] and magnetoelectric (ME),[17–20] each concept offering distinct advantages for specific applications. Among these, magnetoelectric sensors exploit the mechanical coupling between magnetostrictive and piezoelectric layers, allowing direct conversion of the magnetic field into electrical signals.[21,22] Their compatibility with micro-electromechanical system (MEMS) fabrication enables miniaturization to the micrometer scale, integration with electronics, and the development of sensor arrays.[23–25] Utilizing the direct magnetoelectric effect allows for achieving detection limits within the low picotesla range.[26,27] Achieving these low detection limits is restricted to a narrow bandwidth around the sensor's resonance frequency.[28,29] To extend the frequency range, various modulation techniques have been employed, including magnetic frequency conversion,[30,31] electrical modulation,[32,33] and the delta-E effect.[34]

The delta-E effect describes the change in the mechanical stiffness tensor of a ferromagnetic material in response to an applied magnetic field.[35,36] The altered stiffness components detune the resonance frequency, resulting in a change in the output signal.[34,37] Such sensors can feature diverse geometries and sizes, including mm-sized cantilevers and surface acoustic wave devices,[38–46] sub-mm-sized plates,[47,48] and double-wing resonators.[49,50] These sensors have been extensively investigated both theoretically and experimentally, with studies evaluating various aspects, including the sensors' sensitivity,[38,39,49] signal and noise behavior,[44,46,50–52] frequency response,[42] and the influence of resonance modes,[39,45,49] piezoelectric material,[53] and quality factors[40] on their performance.

Despite this extensive work, most studies on delta-E effect sensors focus on individual sensor elements that serve as model systems for specific, sometimes isolated aspects of these devices. However, one of the key challenges in tailoring delta-E effect sensors lies in understanding how performance characteristics such as signal and noise depend on design aspects that have complex secondary effects on the performance by influencing other device parameters. For example, magnetic domain structure and behavior, internal stress distributions, and magnetic anisotropy strongly influence sensor performance and are all affected by fundamental structural design aspects such as resonator geometry and magnetic layer thickness.[38,48] Directly connecting these aspect with sensor performance through physically accurate models, e.g., of the magnetic microstructure and its spatio-temporal evolution, remains computational prohibitive on realistic scales,[54] owing to micrometer device geometries and the low-frequency signals of interest.

Statistical experimental investigations of signal and noise could be a way of identifying empirical connections between design aspects and performance to develop reliable design guidelines. For example, understanding the contribution of the individual noise sources permits developing targeted strategies to mitigate specific noise mechanisms without compromising sensitivity. Identifying how geometry influences the balance between sensitivity and noise is essential for determining the limits of miniaturization for individual devices and, particularly, more complex, multi-element sensor arrays. These insights help define how far delta-E effect sensors can be scaled down without compromising performance and provide guidance for designing next-generation sensor systems.

Here, we present a systematic analysis of sub-mm-sized double-wing delta-E effect magnetic field sensors across 24 configurations, varying in-plane geometry and magnetic layer thickness. First, we conduct a magnetooptical analysis to examine the magnetic properties.

Second, we analyze the dependency of admittance characteristics and magnetic and electrical sensitivities on magnetic layer thickness for the third resonance mode (RM3). Third, we evaluate the signal, noise, and limit of detection ($LoD$) as functions of magnetic volume, excitation voltage amplitude, and magnetic signal frequency. Lastly, building on the observed results, we identify how critical parameters influence sensor performance and propose design strategies to optimize $LoD$.

## 2. Results
### 2.1. Sensor Concept

The magnetoelectric sensors analyzed in this study are based on a double-wing poly-Si resonator with a thickness of 10 μm, anchored at the center of their long axis (**Figure 1**a). A 0.5-μm-thick AlN piezoelectric layer and 1-μm-thick Al electrodes are deposited on the top surface of the resonator. An amorphous magnetostrictive layer $(Fe_{90}Co_{10})_{78}Si_{12}B_{10}$ (FeCoSiB) with thicknesses $t_\mathrm{m}$ = 100, 200, 400, and 600 nm is deposited on the rear side of the resonator using a shadow mask under a magnetic field applied along the sensors' short axis to induce uniaxial magnetic anisotropy. A schematic cross-section of a resonator is shown in Figure 1b. Details of the fabrication process can be found in Ref. [49]. Six different resonator dimensions are investigated for each FeCoSiB thickness, leading to a total number of 24 sensor elements. The mode shape of the third resonance mode (RM3), which is used throughout this work, is illustrated in Figure 1c. The in-plane dimensions of the sensors and their corresponding third-mode resonance frequencies, ranging from approximately 340 kHz to 1.1 MHz depending on the geometry, at zero magnetic bias are summarized in **Table 1**.

During delta-E operation, a sinusoidal voltage $u_\mathrm{ex}(t)$ ($t$: time) with a frequency $f_\mathrm{ex}$ and amplitude $\hat{u}_\mathrm{ex}$ is applied to the piezoelectric layer via one of the top electrodes and the rear side electrode ator near the sensor's mechanical resonance frequency $f_\mathrm{r}$. An alternating magnetic flux density $B_\mathrm{ac}$ leads to a change in the resonance frequency $f_\mathrm{r}$ via the delta-E effect. This change in the $f_r$ alters the admittance $Y(f_{ex})$, modulating the current through the sensor, which is then read out as an output voltage $u_{co}(t)$ with a charge amplifier. In general, both amplitude and phase modulation can occur. Under the small-signal approximation and for $f_{ex} = f_r$, phase modulation can be neglected. Accordingly, $u_{co}(t)$ can be expressed as,[41]

$$u_{co}(t) \approx |Z_f(f_{ex})| \cdot \hat{u}_{ex} \cdot [Y_0 + S_{am}B_{ac}(t)] \cdot \cos(2\pi f_{ex} t). \quad (1)$$

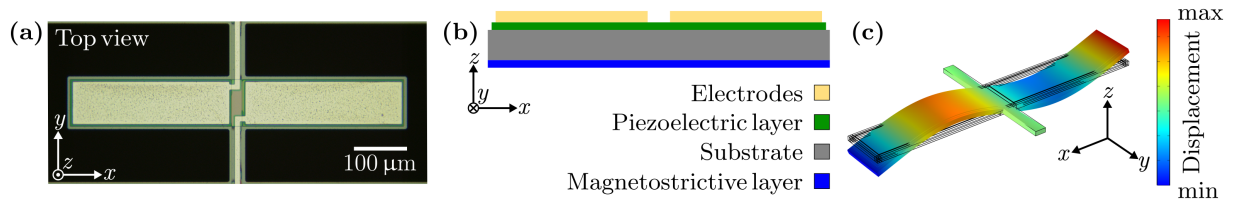

**Figure 1.** (a) Optical microscopy image of the top of an example resonator. (b) Cross-section of a resonator. (c) Mode shape of the third resonance mode (RM3), simulated with a finite element mechanics model.

**Table 1.** In-plane dimensions of the investigated sensors and measured resonance frequencies $f_r$ at third resonance mode (RM3) for FeCoSiB thicknesses of $t_m = 100, 200, 400,$ and $600$ nm at zero magnetic bias flux density $B$.

| Sensor ID | In-plane dimensions [μm × μm] | Resonance frequency $f_r$ [kHz] | | | |
|---|---|---|---|---|---|
| | | 100 nm | 200 nm | 400 nm | 600 nm |
| 1 | 500 × 80 | 1147.7 | 1125.5 | 1138.7 | 1126.9 |
| 2 | 640 × 100 | 699.1 | 688.7 | 698.6 | 691.9 |
| 3 | 740 × 115 | 521.5 | 510.1 | 524.6 | 516.5 |
| 4 | 900 × 120 | 353.1 | 344.5 | 348.4 | 343.9 |
| 5 | 900 × 150 | 349.2 | 345.8 | 347.2 | 345.4 |
| 6 | 900 × 200 | 346.6 | 345.3 | 345.4 | 344.2 |

In this equation, $|Z_f(f_{ex})|$ denotes the feedback impedance magnitude of the charge amplifier, $Y_0 := |Y(f_{ex}, B_0)|$ corresponds to the magnitude of the electrical admittance at $f_{ex}$. A DC magnetic bias flux density $B_0$ is applied along the sensor's long axis during operation. The factor $S_{am}$ represents the total amplitude sensitivity and quantifies how strongly the sensor's admittance responds to changes in magnetic field (see Experimental Section for full definition). It is influenced by magnetic, mechanical and electrical properties of the sensor, which are analyzed in detail in the Electromechanical Properties section.

## 2.2. Magnetic Properties

The magnetic properties of the FeCoSiB layer play a significant role in determining the frequency response, sensitivity, and noise of delta-E effect sensors. Since both magnetic layer thickness and in-plane geometry can influence magnetization behavior, susceptibility, and magnetic domain configuration, characterizing these dependencies is important for understanding and optimizing sensor performance. The magnetic properties of the sensors are investigated using magnetooptical Kerr effect (MOKE) microscopy.[55] Example magnetization curves of sensor ID 2 with FeCoSiB layer thicknesses of $t_m = 100, 200, 400,$ and $600$ nm are shown in **Figure 2**a-d. For the measurements, the magnetic field was applied along the long axis of the resonators, with the magnetooptical sensitivity aligned in the same direction.

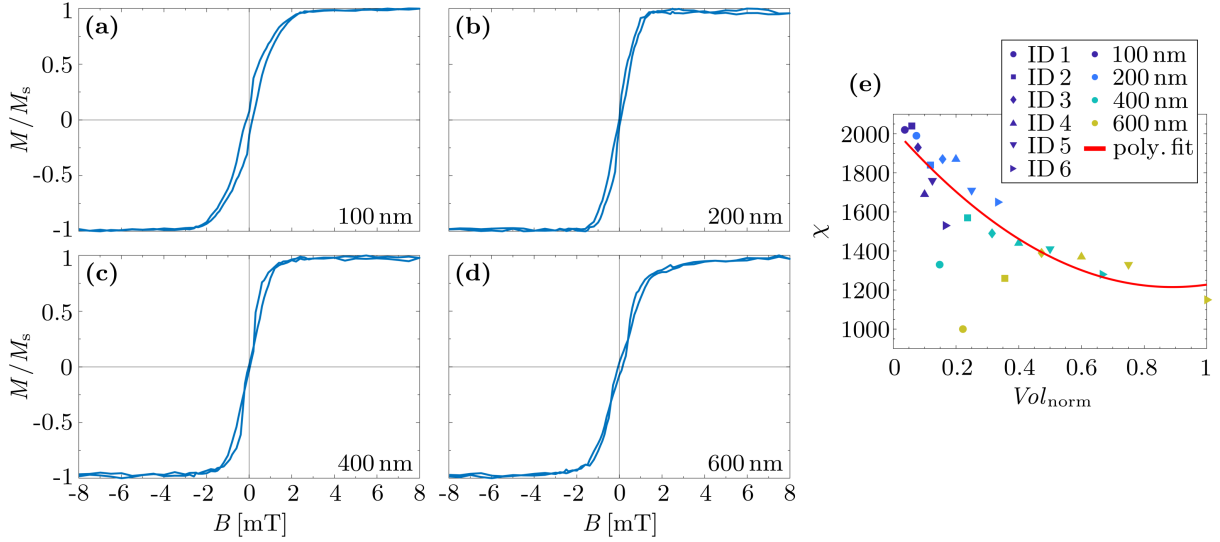

**Figure 2.** (a-d) Magnetization curves of sensor ID 2 for FeCoSiB layer thicknesses $t_\mathrm{m} = 100$, 200, 400, and 600 nm with magnetic flux density $B$ applied along the long axis of the resonator. The MOKE sensitivity was aligned parallel to $B$. (e) Differential magnetic susceptibility $\chi$ at $B = 0$ mT measured at the center of the resonators as a function of the normalized magnetic volume $Vol_\mathrm{norm} \coloneqq Vol_\mathrm{mag}/Vol_\mathrm{mag,max}$ for all investigated sensors.

As the thickness increases, the slope of the magnetization curves around $B = 0$ mT decreases. To quantify this trend, the differential magnetic susceptibility $\chi$ around $B = 0$ mT is estimated and plotted in **Figure 2**e as a function of normalized magnetic volume $Vol_\mathrm{norm} \coloneqq Vol_\mathrm{mag}/Vol_\mathrm{mag,max}$, where $Vol_\mathrm{mag}$ is the magnetic volume of each sensor element. A clear decreasing trend is observed. A second-order polynomial fit is added to the figure as a guide to the eye. The decrease in $\chi$ can be attributed to stress-induced anisotropy, due to the accumulated internal stress occurring during layer deposition.[48,49] Under identical deposition conditions, thicker magnetic films may accumulate greater residual stress, leading to larger stress anisotropy and, consequently, a reduction in susceptibility $\chi$. The contribution of shape anisotropy due to demagnetizing fields is comparatively small in the investigated geometries and does not account for the observed changes in susceptibility (Supplementary Information).

## 2.3. Electromechanical Properties

To evaluate the impact of magnetic layer thickness and sensor geometry on frequency response and magnetic and electrical sensitivities, the electromechanical behavior of the sensors is analyzed. Admittance magnitudes $|Y|$ are measured as a function of excitation frequency $f_\mathrm{ex}$ at various magnetic flux densities $B$ applied along the long axis of the sensors between -20 mT and 20 mT, starting from negative magnetic saturation. The measurements are performed at an excitation voltage amplitude of $\hat{u}_{ex} = 10$ mV. A linear modified Butterworth-van-Dyke (mBvD) equivalent circuit model is fitted to the measured data to extract the resonance frequencies $f_r$, quality factors $Q$, and circuit parameters.[56]

**Figure 3**a shows the measured normalized resonance frequencies $f_r/f_{r,max}$ as a function of the magnetic flux density $B$ for FeCoSiB thicknesses of $t_m = 100, 200, 400$, and 600 nm at the third resonance mode (RM3) of sensor ID 2 at $\hat{u}_{ex} = 10$ mV. The normalized frequency detuning $\Delta f_{r,norm} \coloneqq (f_{r,max} - f_{r,min})/f_{r,max}$, which quantifies the difference between the maximum $f_{r,max}$ and minimum $f_{r,min}$ resonance frequencies, increases with FeCoSiB thickness. This behavior reflects the influence of larger magnetic volume fractions. The normalized resonance frequency curves become increasingly asymmetric with rising FeCoSiB thickness. This growing asymmetry correlates with the appearance of the blocked narrow magnetic domain state[57,58] visible as openings of the magnetization curves in Figure 2c,d (additional data is provided in Supplementary Information). A similar asymmetry in the resonance frequency curves have been previously observed and explained for the mm-sized sensors.[59] The $f_r/f_{r,max}(B)$ curves broaden and reach saturation at higher magnetic flux densities with increasing thickness, which can be attributed to the decrease in the magnetic susceptibility $\chi$ discussed earlier.

An important parameter for evaluating the delta-E effect sensor performance is its sensitivity to magnetic fields. To quantify it, the amplitude sensitivity $S_{am} \coloneqq S_{m,r} \cdot S_{el,r}$ is used, defined as the product of the relative magnetic sensitivity $S_{m,r}$ and the relative electrical sensitivity $S_{el,r}$.[39] The relative magnetic sensitivity can be decomposed into two contributions: a mechanical term $\partial_C f_{r,ij}$, which describes the sensitivity of the resonance frequency to changes in the stiffness tensor component $C_{ij}$, and a magnetoelastic term $\partial_B C_{ij}$, which captures how the respective stiffness tensor component changes with applied magnetic flux density via the delta-E effect.[39] As the sensors operate in bending mode, the stiffness component $C_{11}$ is most relevant and is used for evaluating magnetic sensitivity. The details about these quantities are described in the Experimental Section.

Figure 3b shows the modeled mechanical contribution $\partial_C f_{r,11}$. It increases linearly with magnetic layer thickness but remains nearly constant across different geometries for a given thickness. This trend can be explained by the fact that as the magnetic layer becomes thicker relative to the substrate, its influence on the composite stiffness of the structure increases.[38] However, because all geometries share the same layer thickness ratios and substrate thickness, their mechanical stiffness scaling behaves similarly, resulting in almost identical $\partial_C f_{r,11}$ values for different geometries at the same thickness. In contrast, the magnetic contribution shown in Figure 3c as the field sensitivity of the stiffness tensor $\partial_B C_{11}$ decreases with increasing thickness. This parameter is simulated using the macrospin-based model described in Experimental Section, incorporating a second-order polynomial fit to the experimental susceptibility $\chi$ data from Figure 2 to account for the dependence on magnetic layer thickness. The observed reduction in $\partial_B C_{11}$ is consistent with the drop in magnetic susceptibility $\chi$ and is primarily attributed to increased stress-induced anisotropy in thicker FeCoSiB films. Within the model, this thickness-dependent decrease in susceptibility directly reduces the delta-E effect.

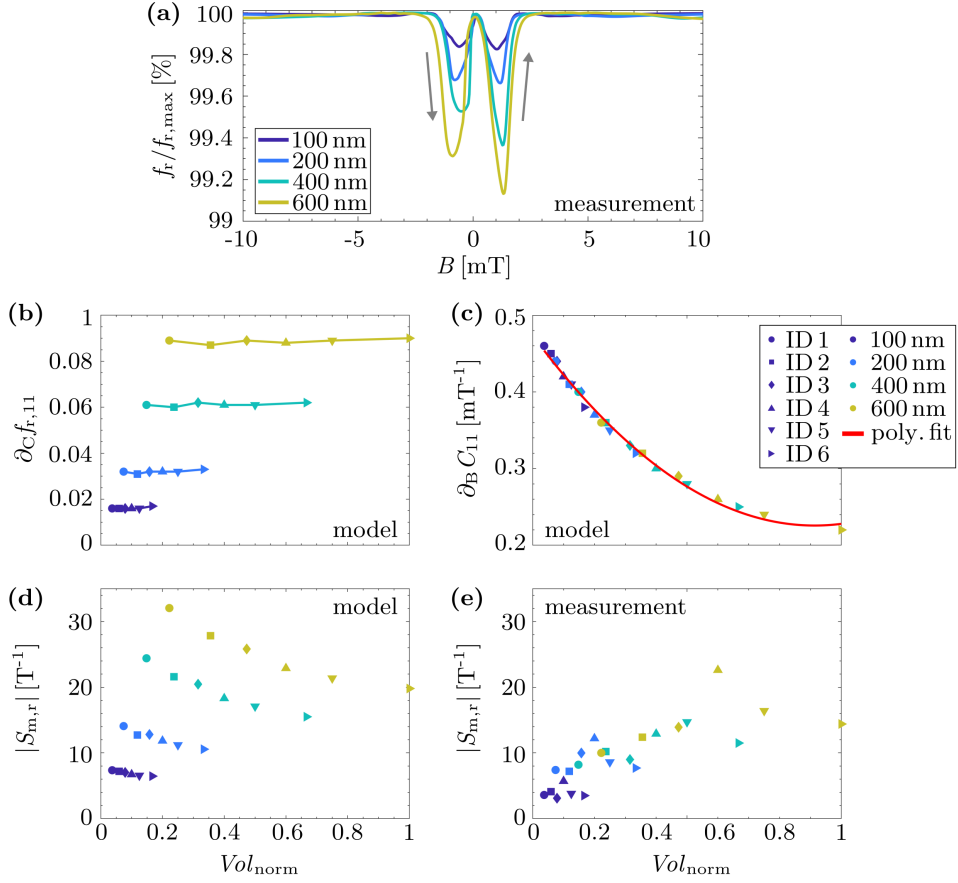

**Figure 3.** (a) Measured normalized resonance frequencies $f_r/f_{r,max}$ as a function of the magnetic flux density $B$ applied along the long axis of the resonator for FeCoSiB thicknesses $t_m = 100, 200, 400$, and $600$ nm at the third resonance mode (RM3) of sensor ID 2 at an excitation voltage amplitude $\hat{u}_{ex} = 10$ mV. Arrows indicate the sweep direction of $B$. (b) Modeled normalized frequency factor $\partial_C f_{r,11}$ as a function of normalized magnetic volume $Vol_{norm} \coloneqq Vol_{mag}/Vol_{mag,max}$, representing the influence of the stiffness tensor component $C_{11}$ on the resonance frequency. (c) Modeled magnetic field sensitivity $\partial_B C_{11}$ of the stiffness tensor component $C_{11}$, evaluated at the inflection point of the inner slope of the positive side of the $C_{11}(B)$ curves using a magnetic model with magnetic susceptibility $\chi$. The red curve shows a second-order polynomial fit. (d) Modeled relative magnetic sensitivity $S_{m,r}$, calculated as the product of $\partial_C f_{r,11}$ and $\partial_B C_{11}$, shown as a function of $Vol_{norm}$. (e) Measured relative magnetic sensitivity $S_{m,r}$ at magnetic flux densities $B_{inf}$ corresponding to the inflection point of the inner slope on the positive side of the $f_r(B)$ curves as a function of normalized magnetic volume $Vol_{norm}$ obtained from experimental data at $\hat{u}_{ex} = 10$ mV.

These opposing trends, an increasing mechanical contribution (Figure 3b) and a decreasing magnetic contribution (Figure 3c), lead to a sublinear increase in the modeled total relative magnetic sensitivity $S_{m,r}$ with magnetic layer thickness, shown in Figure 3d. While $\partial_C f_{r,11}$ remains nearly constant across different geometries for a given thickness, $\partial_B C_{11}$ decreases as the sensor size increases. Consequently, when the magnetic layer thickness is kept constant, but the sensor geometry becomes larger, the reduced magnetic contribution dominates, leading to a slight decrease in $S_{m,r}$ with increasing sensor size. Figure 3e shows the measured relative magnetic sensitivity $S_{m,r}$, which, while following a generally similar increasing trend with magnetic layer thickness as seen in the model (Figure 3d), exhibits slightly lower absolute values and differences across geometries owing to the simplified model approximations.

In addition to magnetic sensitivity, the relative electrical sensitivity $S_{el,r}$ decreases with increasing magnetic layer thickness, as shown in **Figure 4a**. This decline can be attributed to the nonlinear magneto-mechanical losses in the magnetostrictive films.[60] These nonlinear losses increase with the increasing stress amplitude in the magnetostrictive layer resulting in the decrease of the effective quality factor $Q$ of the resonator with $\hat{u}_{ex}$. The corresponding data at two different excitation voltage amplitudes, $\hat{u}_{ex} = 10$ mV and $\hat{u}_{ex} = 300$ mV are shown in Figure 4b-c. A general decline in $Q$ with increasing normalized magnetic volume $Vol_{norm}$ also visible in Figure 4b-c can be explained by larger mechanical losses in the magnetostrictive layer in comparison to the non-magnetic layers.

To test this hypothesis, a numerical finite element mechanics model is used. The model incorporates the resonator with two separate mechanical damping factors: one for the magnetic layer ($\gamma_m$) and one for the non-magnetic components ($\gamma_{nm}$). It assumes that $\gamma_{nm}$ is independent of $\hat{u}_{ex}$ and $\gamma_{nm} < \gamma_m$ to reflect the higher intrinsic losses in the magnetic layer. Both damping factors are assumed to be independent of the magnetic volume. $\gamma_m$ is assumed to be a function of $\hat{u}_{ex}$. Based on these assumptions, the model reproduces the measured trends in the quality factor $Q$ very well confirming the explanation provided above. At 300 mV, the damping factor for the magnetic layer $\gamma_m$ is found to be increased by a factor of approximately 2.5 compared to the value used at 10 mV indicating that the additional losses stem primarily from the magnetic layer and become significantly larger at higher excitation voltage amplitudes.

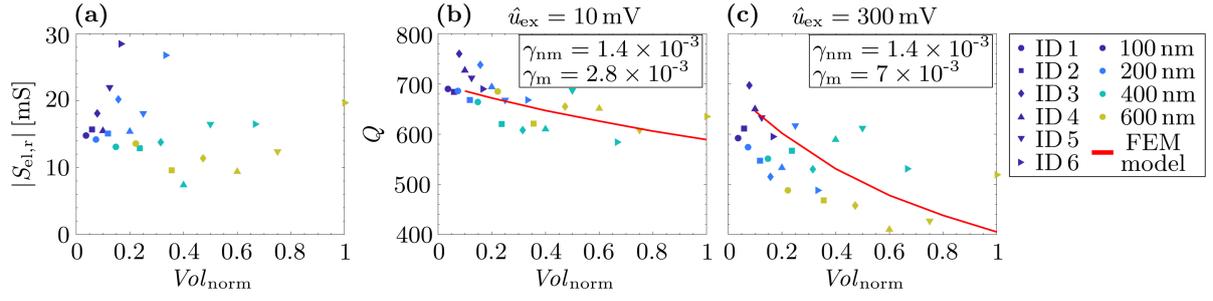

**Figure 4.** (a) Relative electrical sensitivity $S_{el,r}$ measured at an excitation voltage amplitude of $\hat{u}_{ex} = 10$ mV, and (b, c) quality factor $Q$ at magnetic flux densities $B_{inf}$ corresponding to the inflection point of the inner slope on the positive side of the $f_r(B)$ curves as functions of normalized magnetic volume $Vol_{norm}$ for all investigated sensors. Quality factors are shown for excitation voltage amplitudes of (b) $\hat{u}_{ex} = 10$ mV and (c) $\hat{u}_{ex} = 300$ mV. The red curves represent simulations based on a finite element model. Separate damping factors are used for magnetic and non-magnetic layers ($\gamma_m, \gamma_{nm}$). Values used in the simulations are indicated.

## 2.4. Signal and Noise

The capability of detecting weak magnetic fields is not solely determined by sensitivity, but also by the noise level, which ultimately determines the minimum detectable magnetic field - the limit of detection $LoD$. The signal and noise performance is analyzed for different FeCoSiB layer thicknesses (sensor ID 2, $t_m = 100, 200, 400$ and $600$ nm) as a function of excitation voltage amplitude $\hat{u}_{ex}$. **Figure 5**a-f shows the measured and simulated voltage sensitivity $S_V$, amplitude spectral density $E_{co}$, and limit of detection $LoD$ as functions of $\hat{u}_{ex}$ of two example devices (sensor ID 2, $t_m = 100, 600$ nm). The other measurements for $t_m = 200, 400$ nm are provided in the Supporting Information (Figure S5), and the definitions of $S_V$, $E_{co}$, and $LoD$ are provided in the Experimental Section. The signal and noise simulations are based on the model described in Ref. [59]. The output signal is computed using Equation (1) with constant values for the magnetic and electric sensitivity.

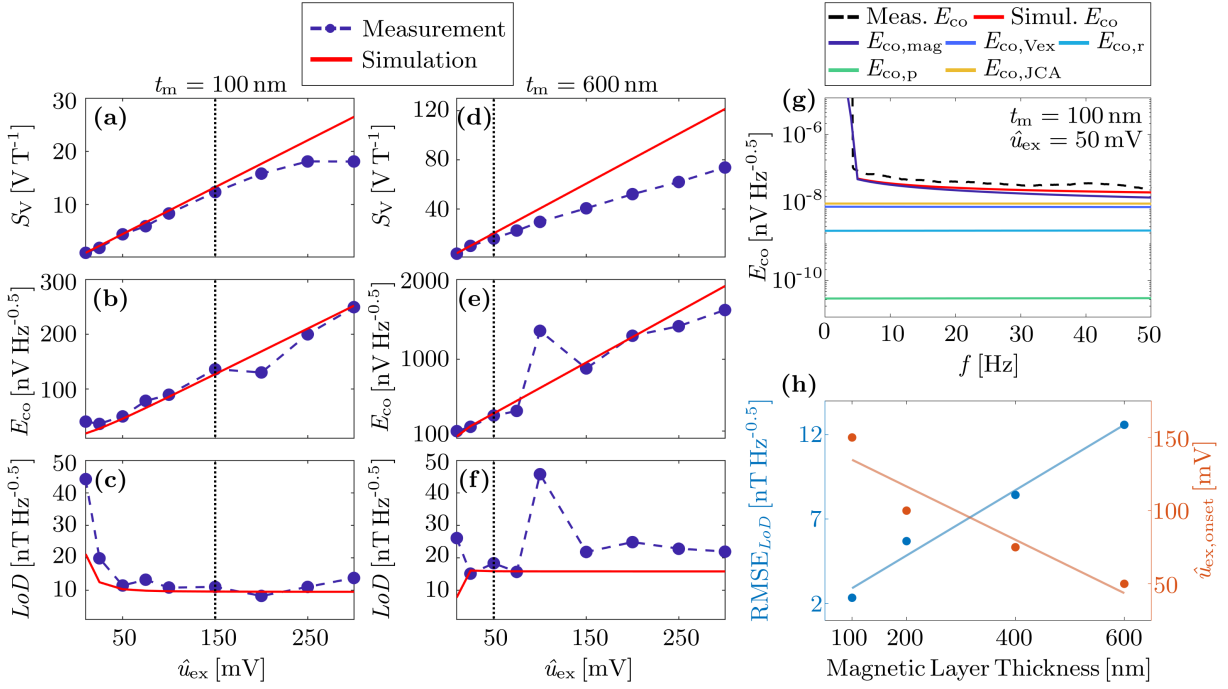

**Figure 5.** Measured and simulated (a, d) voltage sensitivity $S_V$, (b, e) total noise amplitude spectral density $E_{co}$, and (c, f) limit of detection $LoD$ for a magnetic signal with an amplitude of $\hat{B}_{ac} = 1\,\mu T$ and a frequency of $f_{ac} = 10\,Hz$, plotted as functions of excitation voltage amplitude $\hat{u}_{ex}$. Measurements and simulations are conducted at magnetic flux densities $B_{inf}$ corresponding to the inflection point of the inner slope on the positive side of the $f_r(B)$ for a FeCoSiB thicknesses of (a-c) $t_m = 100\,nm$ and (d-f) $t_m = 600\,nm$ at the third resonance mode RM3 of sensor ID 2. The black dotted vertical lines indicate the excitation voltage amplitude $\hat{u}_{ex,onset}$, corresponding to the onset of nonlinearities. (g) Comparison of measured and simulated amplitude spectral density $E_{co}$ at a magnetic flux density of $B_0 = 0.42\,mT$ for a FeCoSiB thickness of $t_m = 100\,nm$ at RM3 of sensor ID 2, with an excitation voltage amplitude of $\hat{u}_{ex} = 50\,mV$. The model includes carrier-mediated magnetic noise density $E_{co,mag}$, the excitation signal's noise density $E_{co,Vex}$, thermal-mechanical noise density $E_{co,r}$, thermal-electrical noise density $E_{co,p}$ of the piezoelectric layer, and the charge amplifier's noise density $E_{co,JCA}$. (h) Root mean square error $RMSE_{LoD}$ between measured and simulated limit of detection $LoD$ and onset excitation voltage amplitude $\hat{u}_{ex,onset}$ where nonlinearities arise, plotted against magnetic layer thicknesses. Results for $RMSE_{LoD}$ are evaluated for $\hat{u}_{ex} = 50\text{-}300\,mV$, where the magnetic noise density $E_{co,mag}$ is expected to be the dominant noise source. Linear fits are shown as a guide to the eye.

At low excitation voltages, all sensors exhibit a linear increase in voltage sensitivity $S_V$, with measurements closely matching simulations. This linear trend is consistent with the constant magnetic and electrical sensitivities used for simulations. With increasing excitation amplitude $\hat{u}_{ex}$, this trend becomes sublinear, and for some thicknesses, even saturates or declines at higher voltages (e.g., 100 nm, Figure 5a). As the magnetic layer thickness increases, the deviations between simulation and measurement appear at progressively lower excitation voltage amplitudes. This trend is evident when comparing the 100-nm-thick and 600-nm-thick FeCoSiB layers in Figure 5a, d, and is quantitatively illustrated in Figure 5h, which shows that the onset excitation voltage amplitude $\hat{u}_{ex,onset}$ decreases with increasing $t_m$. These deviations from linearity can be attributed to the onset of magnetostriction-induced resonator nonlinearities discussed above that reduce the quality factor and electrical sensitivity $S_{el,r}$ with increasing excitation amplitude and magnetic layer thickness. This interpretation is consistent with previous experiments on the dependency of $S_V$ on $\hat{u}_{ex}$ for fixed magnetic layer thickness,[53,59,61] and in line with our thickness-dependent analysis of the electrical sensitivity (Supplementary Information, Figure S3) and the quality factor (Figure 4).

The noise model assumes linear sensor behavior and includes various noise sources contributing to the total amplitude noise density $E_{co}$, such as thermal-mechanical noise $E_{co,r}$, thermal-electrical noise of the piezoelectric layer $E_{co,p}$, charge amplifier's noise $E_{co,JCA}$, and the excitation signal's noise $E_{co,Vex}$. The term $E_{co,mag}$ attributed to the intrinsic magnetic noise is added to account for the deviations between the measured noise level and the simulated non-magnetic noise level. Output-referred magnetic noise $E_{co,mag}$ is assumed to be linearly proportional to the voltage sensitivity $S_v$, similar to how it was introduced in Ref. [59].

At small $\hat{u}_{ex}$ (below 50 mV) and for all magnetic layer thicknesses, the noise floor is dominated by electronic noise sources (charge amplifier's and excitation source's noise), consistent with previous reports.[61] At larger $\hat{u}_{ex}$, the sensor performance is well described by the magnetic noise term. A representative example of the individual noise contributions for a sensor with $t_m$ = 100 nm, operated in this regime ($\hat{u}_{ex}$ = 50 mV), is shown in Figure 5g. At this excitation level, magnetic noise dominates, followed by electronic noise from the charge amplifier and the excitation source. At certain excitation voltage amplitudes, depending on the magnetic layer thickness, the measured noise level begins to fluctuate significantly around the simulated noise level. The onset of these fluctuations correlates with deviations from the linear voltage sensitivity trend; specifically, increasing FeCoSiB thickness results in earlier onset and larger deviations of the simulated $E_{co}$ from the measurements (Figure 5b,e). This behavior is consistent with the earlier onset of the nonlinear operation regime in thicker films.

The limit of detection ($LoD$) is shown in Figure 5c,f as a function of the excitation voltage amplitude $\hat{u}_{ex}$. At low $\hat{u}_{ex}$ (<50 mV), the measured $LoD$ is relatively high across all thicknesses due to the dominance of electronic noise sources and low $S_V$. As $\hat{u}_{ex}$ increases (between 50 and 150 mV in Figure 5c), the $LoD$ remains nearly constant with only minor fluctuations, and the measurements align well with the simulations. This indicates that magnetic noise dominates in this range and is accurately captured by the model. When $\hat{u}_{ex}$ is increased further, the deviation between simulated and measured $LoD$ becomes more pronounced, marking the onset of the nonlinear regime. Figure 5h quantifies this behavior by showing the root mean square error between simulated and measured $LoD$ values ($\text{RMSE}_{LoD}$) in the $\hat{u}_{ex}$ range of 50-300 mV. $\text{RMSE}_{LoD}$ increases with magnetic layer thickness and correlates with the decreasing excitation voltage $\hat{u}_{ex,onset}$ at which the nonlinear regime begins.

Overall, three distinct operating regimes can be identified for the sensor with $t_m = 100$ nm (Figure 5a-c): dominant non-magnetic noise at low $\hat{u}_{ex}$, dominant magnetic noise at moderate $\hat{u}_{ex}$, and the nonlinear regime at $\hat{u}_{ex} > \hat{u}_{ex,onset}$. Similar regimes have been reported for millimeter-sized sensors.[59] However, our results demonstrate that these regimes are strongly influenced by the magnetic layer thickness. In particular, as the magnetic layer thickness increases, the onset of the nonlinear regime shifts to lower excitation voltages, diminishing the intermediate operating regime. For the sensor with $t_m = 600$ nm (Figure 5d-f), this intermediate regime is no longer visible due to the very low $\hat{u}_{ex,onset} < 50$ mV, causing the first and third regimes to overlap.

*2.4.1 Statistical performance analysis*

To statistically assess how sensor performance vary with magnetic volume and excitation voltage amplitude $\hat{u}_{ex}$, the voltage sensitivity $S_V$, amplitude spectral density $E_{co}$ and limit of detection $LoD$ are compared for all investigated sensors as functions of normalized magnetic volume $Vol_{norm}$ with a magnetic signal frequency of $f_{ac} = 10$ Hz at excitation voltage amplitudes of $\hat{u}_{ex} = 10$ mV, 50 mV and 300 mV (**Figure 6**). Measurements for each magnetic layer thickness are conducted at the magnetic bias flux densities $B$ near the inflection points on the inner slope of the resonance frequency curves.

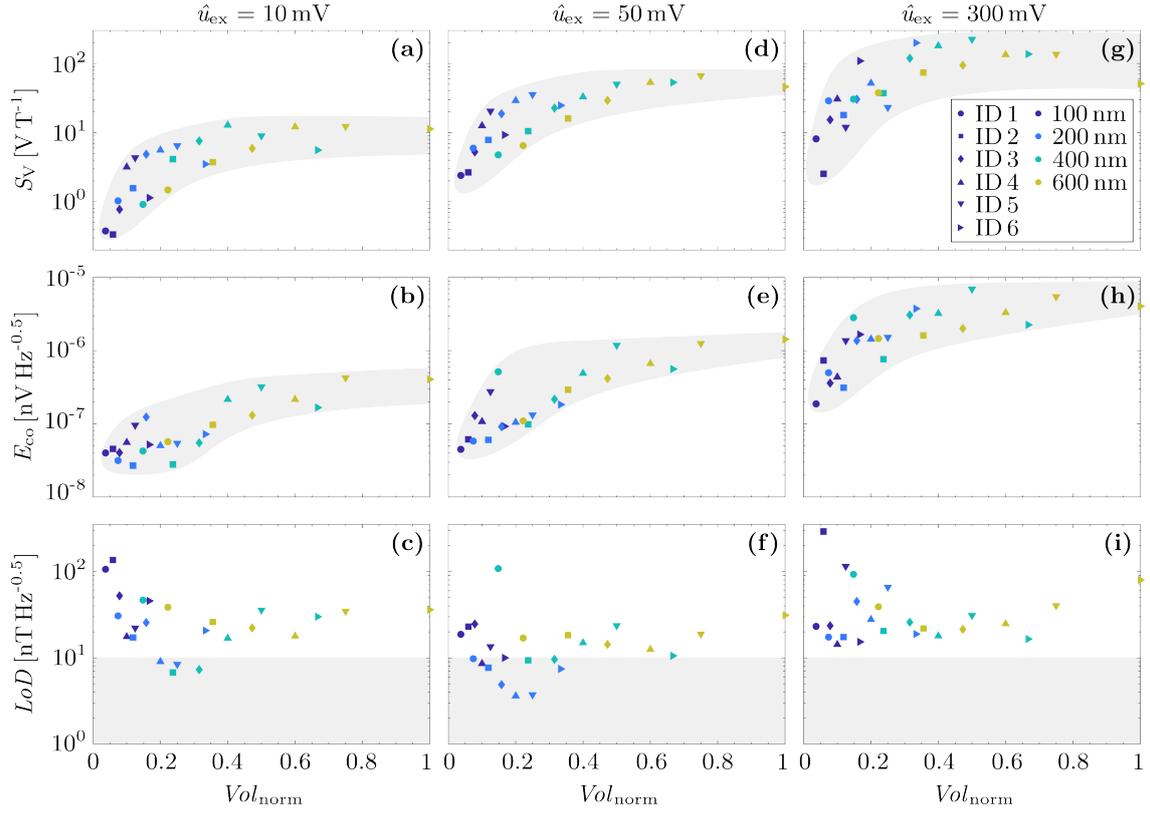

**Figure 6.** Voltage sensitivity $S_V$, amplitude spectral density $E_{co}$, and limit of detection $LoD$ for a magnetic signal with an amplitude of $\hat{B}_{ac} = 1\,\mu\text{T}$ and a frequency of $f_{ac} = 10\,\text{Hz}$ as a function of normalized magnetic volume $Vol_{norm}$ for all investigated sensors at excitation voltage amplitude of $\hat{u}_{ex} = 10\,\text{mV}$ (a-c), $50\,\text{mV}$ (d-f), and $300\,\text{mV}$ (g-i). Gray shaded areas in the $S_V$ (a, d, g) and $E_{co}$ (b, e, h) plots indicate the range of measured values across all sensors, serving as a visual guide. The gray band in the $LoD$ plots (c, f, i) marks the region below $10\,\text{nT}/\sqrt{\text{Hz}}$ to demonstrate the number and distribution of sensors for each magnetic layer thickness achieving $LoD$ values under this threshold at each $\hat{u}_{ex}$.

The voltage sensitivity $S_V$ generally increases with both normalized magnetic volume $Vol_{norm}$ and excitation voltage amplitude $\hat{u}_{ex}$. This trend is expected as magnetic sensitivity generally increases with the magnetic volume (Figure 3e). Sensors with the largest volumes do not always exhibit the highest $S_V$ at high excitation ($\hat{u}_{ex} = 300\,\text{mV}$). Similar behavior and its connection with magnetoelastic nonlinearities are discussed in the previous section for the case of sensor ID 2.

Similar to $S_V$, the amplitude spectral density $E_{co}$ exhibits an increasing trend with both normalized magnetic volume $Vol_{norm}$ and excitation voltage amplitude $\hat{u}_{ex}$. This behavior is primarily caused by the increased magnetic noise and the contribution of the magnetoelastic nonlinearities to the sensor noise, as discussed for sensor ID 2 in Figure 5.

The limit of detection $LoD$ is significantly influenced by excitation voltage amplitude $\hat{u}_{ex}$, FeCoSiB thickness, and sensor geometry. At $\hat{u}_{ex} = 10$ mV, noise levels are lower, but the reduced sensitivity results in higher $LoD$ values. Conversely, at $\hat{u}_{ex} = 300$ mV, sensitivity improves but is counteracted by a more substantial increase in noise, leading to less favorable $LoD$ values. The optimal balance between voltage sensitivity and noise is observed at $\hat{u}_{ex} = 50$ mV, where the $LoD$ is minimized. At $\hat{u}_{ex} = 50$ mV, 10 sensors achieve an $LoD \leq 10$ nT Hz$^{-0.5}$, compared to only 4 sensors at $\hat{u}_{ex} = 10$ mV and none at $\hat{u}_{ex} = 300$ mV. When comparing FeCoSiB thicknesses at $\hat{u}_{ex} = 50$ mV, sensors with a 200 nm layer thickness demonstrate the best $LoD$ performance. While 100 nm sensors show low noise, they suffer from poor sensitivity. In contrast, 400 nm and especially 600 nm sensors provide higher sensitivity but at the cost of elevated noise. Only sensors with 200 nm thickness consistently achieve $LoD \leq 10$ nT Hz$^{-0.5}$, offering the best compromise between sensitivity and noise. Examining sensor geometries, an $LoD \approx 3.6 \pm 1.5$ nT Hz$^{-0.5}$ at $f_{ac} = 10$ Hz at $\hat{u}_{ex} = 50$ mV is achieved by the sensor ID 4 with in-plane dimensions of 900 μm × 120 μm, featuring a 200 nm FeCoSiB layer thickness. Minimum $LoD \approx 800 \pm 110$ pT Hz$^{-0.5}$ at $f_{ac} = 162$ Hz is obtained with the same sensor.

## 3. Discussion

To generalize the findings from the experimental and computational analysis presented in the previous section and visualize the impact of key design parameters, **Figure 7** schematically summarizes the typical behavior of the $LoD$ across three regimes. While similar regimes were previously proposed for mm-sized sensors,[59] our results extend this understanding by revealing how the regime boundaries shift with sensor geometry. This allows for a comprehensive discussion of the complete parameter dependencies governing the $LoD$ performance in miniaturized delta-E effect sensors.

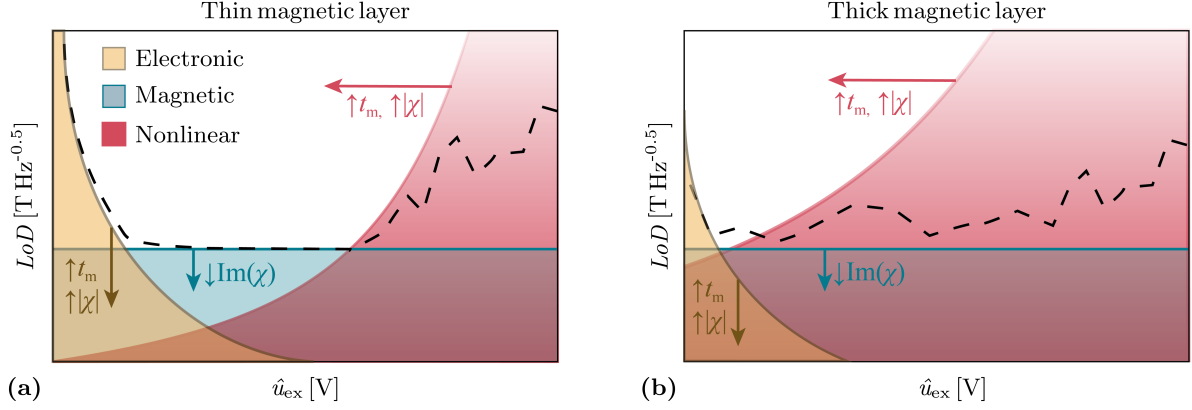

**Figure 7.** Limit of detection *LoD* illustrated as a function of excitation voltage amplitude $\hat{u}_{ex}$, for two examples: (a) thin magnetic layers, and (b) thick magnetic layers. Both examples include three noise regimes: electronic noise dominated (yellow), magnetic noise limited (blue), and nonlinearity dominated (red). Design and material parameters such as magnetic layer thickness $t_m$, absolute magnetic susceptibility $|\chi|$, and its imaginary part $\text{Im}(\chi)$ that encompasses magnetic losses, influence the extent of each region. Up and down arrows next to the parameters indicate whether the denoted trend is connected with the increase (↑) or decrease (↓) of the respective parameter.

The optimal operational point for minimizing *LoD* lies in the magnetic noise-dominated regime (blue region in Figure 7a), where sensitivity is sufficiently high and magnetoelastic nonlinearities have not yet degraded the sensor performance. As indicated, this regime can be expanded or shifted by tuning the thickness $t_m$ and susceptibility $\chi$ of the magnetostrictive layer.

At low excitation voltages and thin magnetic layers (yellow region in Figure 7a), the sensor performance is limited by electronic noise contributions, primarily from the charge amplifier and the excitation source, as consistently observed across all investigated sensor geometries (Section 2.4.). Therefore, this regime can be shifted to smaller excitation voltages and *LoD*s by increasing sensitivity using thicker magnetic layers and higher magnetic susceptibilities. Additionally, electronic noise can be reduced by integrating low-noise charge amplifiers and stable excitation sources tailored for miniaturized sensors.

In the intermediate regime in Figure 7a, where magnetic noise limits performance, the focus should shift to improving effective magnetic material properties. Specifically, increasing magnetic susceptibility $|\chi|$ and decreasing magnetic losses, which are partially encompassed by the imaginary part of $\chi$. The losses are particularly influenced by the number and dynamics of domain walls, their energy, and material defects, is crucial to reduce *LoD*.[59,62] The obvious change of the magnetic domain wall structure and the related effects are not part of this study. Reducing loss and increasing $|\chi|$ can be pursued through magnetic layer and domain engineering, including control of internal stress, reduction of defect-induced pinning sites, and domain wall energy tuning. Using materials or multi-layer designs that promote single-domain behavior (e.g., exchange bias layers) can further improve magnetic sensitivity and reduce noise; therefore, improve $LoD$. Such stacks have been, for example, employed for an alternative ME sensor concept in Ref. [63,64].

At the same time, delaying the onset of nonlinearities to push the nonlinearity-dominated regime (Figure 7, red) to larger excitation voltage amplitudes and extend the optimum operation regime is not straightforward because these nonlinearities are inherently present in magnetostrictive materials. Here, changes in the magnetic domain wall structure with magnetic bias, like the transformations from Bloch to interacting Néel walls might contribute to the results. It is possible to minimize the contribution of the magnetoelastic nonlinearities to the sensor signal by decreasing the magnetic-to-total resonator thickness ratio $t_\mathrm{m}/h$ (Section 2.4.). This moves the nonlinear regime to higher excitation voltages, however, reducing $t_\mathrm{m}/h$ simultaneously decreases the sensor's sensitivity and by that it also extends the electronic-noise-dominated regime to larger excitation voltages. Decreasing the magnetic susceptibility $|\chi|$ is expected to have a similar effect. Hence, to be viable, these strategies may require further reducing electronic noise by employing noise-improved electronics. Consequently, for given operating electronics a clear trade off results: for thin magnetic layers the electronic-noise-limited regime becomes dominant, while for very thick magnetic layers the nonlinear operation regime sets on at smaller excitation voltage amplitudes and extends over a large voltage regime. If this is combined with substantial electronic noise the optimum operation regime vanishes as illustrated in Figure 7b, reflecting the behavior of the devices with 600-nm-thick magnetic layer (Figure 5f).

These findings demonstrate that careful balancing of mechanical, electrical and magnetic contributions is critical in designing miniaturized delta-E effect sensors with both high sensitivity and low noise. The results outline design guidelines for tailoring sensor performance to specific requirements, based on statistical analysis.

## 4. Conclusion

We presented a comprehensive experimental and theoretical study on the influence of magnetic layer thickness and geometry on the performance of sub-mm-sized delta-E effect magnetic field sensors. By evaluating 24 sensor configurations, we investigated magnetic properties, electromechanical behavior, and signal-noise characteristics to identify design strategies that balance sensitivity and noise while minimizing $LoD$.

Magnetic analysis reveals that the magnetic susceptibility decreases with increasing magnetic layer thickness, primarily due to the accumulation of residual stress. Admittance measurements show that thicker magnetic layers lead to enhanced frequency detuning owing to increased magnetic volume fraction but also induce nonlinearities and asymmetry in the resonance behavior. Electromechanical finite element simulations and magnetic modeling indicate that while mechanical contributions to magnetic sensitivity increase with thickness, magnetic contributions decline due to increased effective anisotropy, resulting in only sublinear improvement in magnetic sensitivity. Furthermore, thicker layers reduce electrical sensitivity and quality factor due to increased magneto-mechanical losses, particularly at higher excitation voltage amplitudes, limiting the benefits of larger magnetic volumes.

The voltage sensitivity generally increases with increasing excitation voltage amplitude and magnetic layer thickness, but this trend becomes sublinear at higher excitations due to magnetoelastic nonlinearities. Noise analysis reveals a transition from electronic noise dominance at low excitation voltage amplitudes to magnetic noise at moderate levels, followed by nonlinear behavior at high excitation. Moreover, the onset of nonlinearities occurs earlier as magnetic layer thickness increases, limiting the usable excitation range and degrading the limit of detection $LoD$. A statistical comparison across all sensors confirmed that the best $LoD$ is achieved at intermediate magnetic volumes and moderate excitation voltage amplitudes.

Finally, the results demonstrate the importance of optimized-not maximized-magnetic layer thickness and volume. Excessively thin layers reduce sensitivity, while too thick films lead to early-onset nonlinearities and elevated noise. Effective sensor design requires low-noise electronics, reduced anisotropy through material and domain engineering, and mechanical design strategies to minimize dynamic stress. Compared to previously reported sub-mm-sized delta-E effect sensors with significantly thicker magnetic layers in the literature,[53,65] our sensors with a 200 nm layer thickness and a factor of 10-25 times lower $t_\mathrm{m}/h$ ratio achieve superior $LoD$, demonstrating that compact, low-noise operation is achievable without sacrificing sensitivity. Overall, these findings provide statistically supported design strategies for next-generation miniaturized delta-E effect sensors, enabling performance optimization tailored to diverse application requirements.

## 5. Experimental Section

*Sensor Preparation*: The magnetoelectric MEMS sensors were designed at Kiel University and fabricated through Science Corporation. The microfabrication process, up to the deposition of the magnetic layers, has been described in detail elsewhere.[49,66] The magnetic layers were deposited as the final step at Kiel University to complete the sensor preparation.

FeCoSiB layers with thicknesses of $t_\mathrm{m} = 100, 200, 400$, and $600$ nm sandwiched between 10 nm Ta adhesion and protection layers were deposited on the rear side of the released resonators using DC magnetron sputtering. This process was carried out on four separate chips, each corresponding to one specific FeCoSiB thickness. To ensure precise layer deposition, shadow masks matching the size of the chips were carefully aligned and placed on the rear side of the chips. The magnetic layer deposition was performed in a custom-built deposition chamber at a working pressure of $3 \times 10^{-3}$ mbar with an Ar gas flow of 38 sccm and sputtering power of 20 W. A magnetic field of 130 mT was applied along the short axis of the sensors during deposition using two $Nd_2Fe_{14}B$ permanent magnets to induce a uniaxial magnetic easy axis. Further details regarding the magnetic layer deposition process can be found in previous work.[49]

*Magnetic Characterization:* The magnetic properties of the sensors were investigated using Magnetooptical Kerr effect (MOKE) microscopy.[55] The quasi-static magnetization curves were measured by applying magnetic fields along the long axis of the sensors, while aligning the magnetooptical sensitivity axis in the same direction. Domain images in the demagnetized state were obtained by applying a decaying sinusoidal magnetic field along the long axis of the sensors (Supporting Information). During imaging, the magnetooptical sensitivity axis was aligned along the short axis of the sensors. The images show that the magnetic anisotropy in all samples is well oriented along the short axis of the resonator and any local deviations are expected and result from the local demagnetizing field and anisotropic stress relaxation.[49]

*Magnetic and Electrical Sensitivities:* Admittance magnitudes $|Y|$ were measured as a function of excitation frequency $f_\mathrm{ex}$ and magnetic flux density $B$, using an Agilent 4294A Precision Impedance Analyzer. The amplitude sensitivity $S_\mathrm{am}$ was used to quantify the change of $|Y|$ with $B$. It is expressed as a product of relative magnetic sensitivity $S_\mathrm{m,r}$ and relative electrical sensitivity $S_\mathrm{el,r}$.[39] Sensitivities are normalized to excitation frequency $f_\mathrm{ex} = f_\mathrm{r}$ to be able to compare different sensor thicknesses and geometries operated at varying resonance frequencies. The amplitude sensitivity is given by

$$S_{\text{am}} := \left.\frac{\partial |Y|}{\partial B}\right|_{B=B_0, f=f_r} = S_{\text{m,r}} \cdot S_{\text{el,r}}, \tag{2}$$

where magnetic sensitivity $S_{\text{m,r}}$ and relative electrical sensitivity $S_{\text{el,r}}$ can be defined as

$$S_{\text{m,r}} := \left.\frac{1}{f_r}\frac{\partial f_r}{\partial B}\right|_{B=B_0}, \quad S_{\text{el,r}} := \left.\frac{\partial |Y|}{\partial f}\right|_{f=f_r} \cdot f_r, \tag{3}$$

with the resonance frequency $f_r$, the magnetic bias flux density $B = B_0$ for an operating frequency $f = f_r$. $S_{\text{m,r}}$ can be separated into two parts: a mechanical contribution and a magnetic contribution:[39]

$$S_{\text{m,r}} := \sum_{i=1}^{3}\sum_{j=1}^{3} \partial_C f_{r,ij} \partial_B C_{ij} \tag{4}$$

$$\partial_C f_{r,ij} := \left.\frac{C_{ij}}{f_r}\frac{\partial f_r}{\partial C_{ij}}\right|_{B=B_0}; \quad \partial_B C_{ij} := \left.\frac{1}{C_{ij}}\frac{\partial C_{ij}}{\partial B}\right|_{B=B_0} \tag{5}$$

The mechanical part $\partial_C f_{r,ij}$ represents the sensitivity of the resonance frequency $f_r$ to changes in the stiffness tensor component $C_{ij}$. The magnetic part $\partial_B C_{ij}$, reflects how the stiffness tensor component $C_{ij}$ changes with the magnetic flux density $B$.

*Signal and Noise Measurements:* Signal and noise measurements were carried out using a lock-in amplifier (MFLI, Zurich Instruments). A JFET-based charge amplifier with a feedback capacitance $C_f = 33$ pF and a feedback resistance $R_f = 5$ GΩ was used.[44] Measurements were performed in a magnetically (Model ZG1, Aaronia), electrically, and acoustically shielded setup.[67] Various magnetic bias flux densities were applied along the long axis of the sensors during both signal and noise measurements. Before each measurement, the sensors were saturated by applying a negative magnetic flux density of $B = -20$ mT along the long axis of the sensors using a current source (KEPCO BOP20-10ML), which was then reduced to the desired magnetic bias flux density for the measurement using another low-noise current source (Keysight B2962A). Measurements were performed in two stages: as a function of excitation voltage amplitude $\hat{u}_{\text{ex}}$ and as a function of magnetic signal frequency $f_{\text{ac}}$. For the first set of measurements, excitation voltage amplitude varied incrementally from $\hat{u}_{\text{ex}} = 10$ mV to $\hat{u}_{\text{ex}} = 300$ mV. Signal and noise were measured at the resonance frequency $f_{\text{ex}} = f_r$, which was extracted by fitting modified Butterworth-van-Dyke (mBvD) to the admittance measurements. Noise measurements were performed without a magnetic test signal $B_{\text{ac}}(t)$, while subsequent signal measurements were carried out by applying a sinusoidal magnetic test signal $B_{\text{ac}}(t)$ with a frequency of $f_{\text{ac}} = 10$ Hz and an amplitude of $\hat{B}_{\text{ac}} = 1$ μT using a low-noise current source (Lake Shore M81). The resulting output voltage $u_{co}(t)$ from the sensor was recorded. The second set of measurements was conducted as a function of magnetic signal frequency $f_{ac}$, while keeping the excitation voltage amplitude constant at $\hat{u}_{ex} = 50$ mV. All measurements were performed at a magnetic bias flux density, corresponding to the inflection point on the inner slope of the positive side of the $f_r(B)$ curves.

The voltage amplitude spectrum $U_{RMS}(f)$ of $u_{co}(t)$ was used to calculate the voltage sensitivity $S_V$,

$$S_V(f_{ac}) := \frac{U_{RMS}(f_{ex} + f_{ac})}{\hat{B}_{ac}}. \tag{6}$$

The voltage sensitivity $S_V$ was then used to estimate limit of detection ($LoD$), defined as the smallest detectable magnetic field at a given frequency with one-second averaging time,

$$LoD(f_{ac}) := \frac{E_{co}(f_{ex} + f_{ac})}{S_V(f_{ac})}, \tag{7}$$

where $E_{co}$ is the noise spectral density of $u_{co}(t)$, measured in the absence of a magnetic test signal $B_{ac}$.

*Modeling and Simulation:* A macrospin model was used to estimate the magnetic flux density-dependent stiffness tensor component $C_{11}(B)$. The model assumes uniaxial magnetic anisotropy with the anisotropy axis oriented along the short axis of the sensors and $K_{eff}$ fitted to the slope of the measured magnetization curves around zero magnetic field. The normalized magnetic contribution to relative magnetic sensitivity, $\partial_B C_{11}$, was obtained by numerically differentiating the $C_{11}(B)$ curve at the inflection point of the inner slope on the positive side of the magnetic flux density. Then, finite element simulations were performed in COMSOL Multiphysics v.6.3 to evaluate the change in resonance frequency $f_r$ with respect to stiffness component $C_{11}$, providing the normalized frequency factor $\partial_C f_{r,ij}$. These values were then combined with the magnetic model output to calculate the relative magnetic sensitivity $S_{m,r}$. Additionally, normalized resonance frequency curves as a function of magnetic flux density were simulated for sensor ID 2 across different FeCoSiB layer thicknesses, using the corresponding $K_{eff}$ values.

**Acknowledgements**

The research was funded by the German Research Foundation (Deutsche Forschungsgemeinschaft, DFG) through the Collaborative Research Centre CRC 1261 "Magnetoelectric Sensors: From Composite Materials to Biomagnetic Diagnostics", and the Carl Zeiss Foundation via the project "Memristive Materials for Neuromorphic Electronics" (MemWerk).

**References**

[1] M. Díaz-Michelena, *Sensors* **2009**, *9*, 2271.

[2] C. P. . Treutler, *Sens. Actuators, A* **2001**, *91*, 2.

[3] F. J. Villanueva, D. Villa, M. J. Santofimia, J. Barba, J. C. Lopez, *IEEE World Forum Internet Things, WF-IoT 2015 - Proc.* **2015**, 751.

[4] I. Ashraf, S. Hur, M. Shafiq, Y. Park, *Sensors* **2019**, *19*.

[5] V. Pasku, A. De Angelis, G. De Angelis, D. D. Arumugam, M. Dionigi, P. Carbone, A. Moschitta, D. S. Ricketts, *IEEE Commun. Surv. Tutorials* **2017**, *19*, 2003.

[6] D. Murzin, D. J. Mapps, K. Levada, V. Belyaev, A. Omelyanchik, L. Panina, V. Rodionova, *Sensors* **2020**, *20*, 1569.

[7] G. Lin, D. Makarov, O. G. Schmidt, *Lab Chip* **2017**, *17*, 1884.

[8] S. Zuo, H. Heidari, D. Farina, K. Nazarpour, *Adv. Mater. Technol.* **2020**, *5*.

[9] M. Melzer, J. I. Mönch, D. Makarov, Y. Zabila, G. S. C. Bermúdez, D. Karnaushenko, S. Baunack, F. Bahr, C. Yan, M. Kaltenbrunner, O. G. Schmidt, *Adv. Mater.* **2015**, *27*, 1274.

[10] K. Wu, D. Tonini, S. Liang, R. Saha, V. K. Chugh, J. P. Wang, *ACS Appl. Mater. Interfaces* **2022**, *14*, 9945.

[11] J. E. Davies, J. D. Watts, J. Novotny, D. Huang, P. G. Eames, *Appl. Phys. Lett.* **2021**, *118*, 062401.

[12] M. Wang, Y. Wang, L. Peng, C. Ye, *IEEE Sens. J.* **2019**, *19*, 9610.

[13] T. P. Tomo, S. Somlor, A. Schmitz, L. Jamone, W. Huang, H. Kristanto, S. Sugano, *Sensors* **2016**, *16*.

[14] P. T. Das, H. Nhalil, M. Schultz, S. Amrusi, A. Grosz, L. Klein, *IEEE Sensors Lett.* **2021**, *5*, 1.

[15] D. Karnaushenko, D. D. Karnaushenko, D. Makarov, S. Baunack, R. Schäfer, O. G. Schmidt, *Adv. Mater.* **2015**, *27*, 6582.

[16] K. Mohri, T. Uchiyama, L. V. Panina, M. Yamamoto, K. Bushida, *J. Sensors* **2015**, *2015*, 718069.

[17] J. Gao, Z. Jiang, S. Zhang, Z. Mao, Y. Shen, Z. Chu, *Actuators* **2021**, *10*, 109.

[18] M. Bichurin, R. Petrov, O. Sokolov, V. Leontiev, V. Kuts, D. Kiselev, Y. Wang, *Sensors* **2021**, *21*, 6232.

[19] P. Martins, R. Brito-Pereira, S. Ribeiro, S. Lanceros-Mendez, C. Ribeiro, *Nano Energy* **2024**, *126*, 109569.

[20] C. Dong, X. Liang, (Jingya) Gao Lilyn, H. Chen, Y. He, Y. Wei, M. Zaeimbashi, A. Matyushov, C. Sun, N. X. Sun, *Adv. Electron. Mater.* **2022**, *8*, 2200013.

[21] M. Fiebig, *J. Phys. D. Appl. Phys.* **2005**, *38*, R123.

[22] D. Viehland, M. Wuttig, J. McCord, E. Quandt, *MRS Bull.* **2018**, *43*, 834.

[23] C. Tu, Z. Q. Chu, B. Spetzler, P. Hayes, C. Z. Dong, X. F. Liang, H. H. Chen, Y. F. He, Y. Y. Wei, I. Lisenkov, H. Lin, Y. H. Lin, J. McCord, F. Faupel, E. Quandt, N. X. Sun, *Materials (Basel).* **2019**, *12*, 2259.

[24] H. Li, Z. Zou, Y. Yang, P. Shi, X. Wu, J. Ou-Yang, X. Yang, Y. Zhang, B. Zhu, S. Chen, *IEEE Trans. Magn.* **2020**, *56*, 4000504.

[25] H. Xi, M. C. Lu, Q. X. Yang, Q. M. Zhang, *Sensors Actuators, A Phys.* **2020**, *311*, 112064.

[26] Y. Wang, D. Gray, D. Berry, J. Gao, M. Li, J. Li, D. Viehland, *Adv. Mater.* **2011**, *23*, 4111.

[27] P. Durdaut, S. Salzer, J. Reermann, V. Robisch, P. Hayes, A. Piorra, D. Meyners, E. Quandt, G. Schmidt, R. Knochel, M. Hoft, *IEEE Sens. J.* **2017**, *17*, 2338.

[28] H. Greve, E. Woltermann, R. Jahns, S. Marauska, B. Wagner, R. Knöchel, M. Wuttig, E. Quandt, *Appl. Phys. Lett* **2010**, *97*, 152503.

[29] R. Jahns, A. Piorra, E. Lage, C. Kirchhof, D. Meyners, J. L. Gugat, M. Krantz, M. Gerken, R. Knöchel, E. Quandt, *J. Am. Ceram. Soc.* **2013**, *96*, 1673.

[30] R. Jahns, H. Greve, E. Woltermann, E. Quandt, R. Knöchel, *Sensors Actuators, A Phys.* **2012**, *183*, 16.

[31] V. Röbisch, E. Yarar, N. O. Urs, I. Teliban, R. Knöchel, J. McCord, E. Quandt, D. Meyners, *J. Appl. Phys.* **2015**, *117*.

[32] P. Hayes, S. Salzer, J. Reermann, E. Yarar, V. Röbisch, A. Piorra, D. Meyners, M. Höft, R. Knöchel, G. Schmidt, E. Quandt, *Appl. Phys. Lett.* **2016**, *108*.

[33] P. Hayes, M. Jovičević Klug, S. Toxværd, P. Durdaut, V. Schell, A. Teplyuk, D. Burdin,

A. Winkler, R. Weser, Y. Fetisov, M. Höft, R. Knöchel, J. McCord, E. Quandt, *Sci. Rep.* **2019**, *9*, 1.

[34] J. Reermann, S. Zabel, C. Kirchhof, E. Quandt, F. Faupel, G. Schmidt, *IEEE Sens. J.* **2016**, *16*, 4891.

[35] J. D. Livingston, *Phys. Stat. Sol. A* **1982**, *70*, 591.

[36] E. W. Lee, *Rep. Prog. Phys.* **1955**, *18*, 184.

[37] B. Spetzler, ΔE-Effect Magnetic Field Sensors, Kiel University, **2021**.

[38] B. Spetzler, C. Kirchhof, E. Quandt, J. McCord, F. Faupel, *Phys. Rev. Appl.* **2019**, *12*, 064036.

[39] B. Spetzler, E. V. Golubeva, R. M. Friedrich, S. Zabel, C. Kirchhof, D. Meyners, J. McCord, F. Faupel, *Sensors* **2021**, *21*, 2022.

[40] B. Spetzler, C. Kirchhof, J. Reermann, P. Durdaut, M. Höft, G. Schmidt, E. Quandt, F. Faupel, *Appl. Phys. Lett* **2019**, *114*, 183504.

[41] B. Spetzler, C. Bald, P. Durdaut, J. Reermann, C. Kirchhof, A. Teplyuk, D. Meyners, E. Quandt, M. Höft, G. Schmidt, F. Faupel, *Sci. Rep.* **2021**, *11*, 5269.

[42] B. Spetzler, E. V Golubeva, C. Müller, F. Faupel, *Sensors* **2019**, *19*.

[43] B. Spetzler, P. Wiegand, P. Durdaut, M. Höft, A. Bahr, R. Rieger, F. Faupel, *Sensors* **2021**, *21*, 7594.

[44] P. Durdaut, J. Reermann, S. Zabel, C. Kirchhof, E. Quandt, F. Faupel, G. Schmidt, R. Knöchel, M. Höft, *IEEE Trans. Instrum. Meas.* **2017**, *66*, 2771.

[45] S. Zabel, J. Reermann, S. Fichtner, C. Kirchhof, E. Quandt, B. Wagner, G. Schmidt, F. Faupel, *Appl. Phys. Lett.* **2016**, *108*, 222401.

[46] A. Kittmann, P. Durdaut, S. Zabel, J. Reermann, J. Schmalz, B. Spetzler, D. Meyners, N. X. Sun, J. McCord, M. Gerken, G. Schmidt, M. Höft, R. Knöchel, F. Faupel, E. Quandt, *Sci. Rep.* **2018**, *8*.

[47] T. Nan, Y. Hui, M. Rinaldi, N. X. Sun, *Sci. Rep.* **2013**, *3*, 1985.

[48] A. D. Matyushov, B. Spetzler, M. Zaeimbashi, J. Zhou, Z. Qian, E. V. Golubeva, C. Tu, Y. Guo, B. F. Chen, D. Wang, A. Will-Cole, H. Chen, M. Rinaldi, J. McCord, F. Faupel, N. X. Sun, *Adv. Mater. Technol.* **2021**, *6*, 2100294.

[49] F. Ilgaz, E. Spetzler, P. Wiegand, F. Faupel, R. Rieger, J. McCord, B. Spetzler, *Sci. Rep.* **2024**, *14*, 11075.

[50] P. Wiegand, S. Simmich, F. Ilgaz, F. Faupel, B. Spetzler, R. Rieger, *IEEE Open J. Circuits Syst.* **2024**, *5*, 398.

[51] V. Schell, E. Spetzler, N. Wolff, L. Bumke, L. Kienle, J. McCord, E. Quandt, D. Meyners, *Sci. Rep.* **2023**, *13*.

[52] S. Zabel, C. Kirchhof, E. Yarar, D. Meyners, E. Quandt, F. Faupel, *Appl. Phys. Lett.* **2015**, *107*, 152402.

[53] B. Spetzler, J. Su, R. M. Friedrich, F. Niekiel, S. Fichtner, F. Lofink, F. Faupel, *APL Mater.* **2021**, *9*, 031108.

[54] A. Hubert, R. Schäfer, *Magnetic Domains: The Analysis of Magnetic Microstructures*, Springer Berlin, Heidelberg, **2009**.

[55] J. McCord, *J. Phys. D. Appl. Phys.* **2015**, *48*.

[56] K. S. Van Dyke, *Proc. Inst. Radio Eng.* **1928**, *16*, 742.

[57] S. Middelhoek, *IBM J. Res. Dev.* **1962**, *6*, 394.

[58] J. Steiner, R. Schäfer, H. Wieczoreck, J. McCord, F. Otto, *Phys. Rev. B* **2012**, *85*, 104407.

[59] E. Spetzler, B. Spetzler, J. McCord, *Adv. Funct. Mater.* **2023**, *34*, 2309867.

[60] E. Spetzler, B. Spetzler, D. Seidler, J. Arbustini, L. Thormählen, E. Elzenheimer, M. Höft, A. Bahr, D. Meyners, J. McCord, *Adv. Sens. Res.* **2024**, 2400109.

[61] F. Ilgaz, E. Spetzler, P. Wiegand, F. Faupel, R. Rieger, J. McCord, B. Spetzler, *Appl. Phys. Lett.* **2025**, *126*, 84103.

[62] N. O. Urs, E. Golubeva, V. Röbisch, S. Toxvaerd, S. Deldar, R. Knöchel, M. Höft, E. Quandt, D. Meyners, J. McCord, *Phys. Rev. Appl.* **2020**, *13*, 1.

[63] L. Thormählen, P. Hayes, E. Elzenheimer, E. Spetzler, G. Schmidt, M. Höft, J. McCord, D. Meyners, E. Quandt, *Appl. Phys. Lett.* **2024**, *124*.

[64] M. Jovičević Klug, L. Thormählen, V. Röbisch, S. D. Toxværd, M. Höft, R. Knöchel, E. Quandt, D. Meyners, J. McCord, *Appl. Phys. Lett.* **2019**, *114*.

[65] R. Jahns, S. Zabel, S. Marauska, B. Gojdka, B. Wagner, R. Knöchel, R. Adelung, F. Faupel, *Appl. Phys. Lett.* **2014**, *105*, 2011.

[66] A. Cowen, G. Hames, K. Glukh, B. Hardy, *PiezoMUMPs(TM) Design Handbook*, **2014**.

[67] R. Jahns, R. Knochel, H. Greve, E. Woltermann, E. Lage, E. Quandt, *MeMeA 2011 - 2011 IEEE Int. Symp. Med. Meas. Appl. Proc.* **2011**, 107.

# Supporting Information

**Operation Regimes and Design Principles of Delta-E Effect Sensors**

*Fatih Ilgaz, Elizaveta Spetzler\*, Patrick Wiegand, Robert Rieger, Jeffrey McCord, Benjamin Spetzler\**

**Magnetic Properties**

Magnetic properties are investigated using magnetooptical Kerr effect (MOKE) microscopy for sensors with FeCoSiB layer thicknesses of 100, 200, 400, and 600 nm. **Figure S1** presents the magnetic domain structures of all sensors in their demagnetized state after applying a decaying sinusoidal magnetic field along the sensors' long axis. The magnetic domains are predominantly aligned along the short axis across all sensors. With increasing FeCoSiB thickness, the number of domains grows, and the domain widths become narrower. **Figure S2** presents the magnetization curves of the investigated sensors (ID 1 and ID 3-6), measured along their long axes. As the magnetic layer thickness increases, the slope of the curves decreases, consistent with the trend observed for ID 2 in the main text.

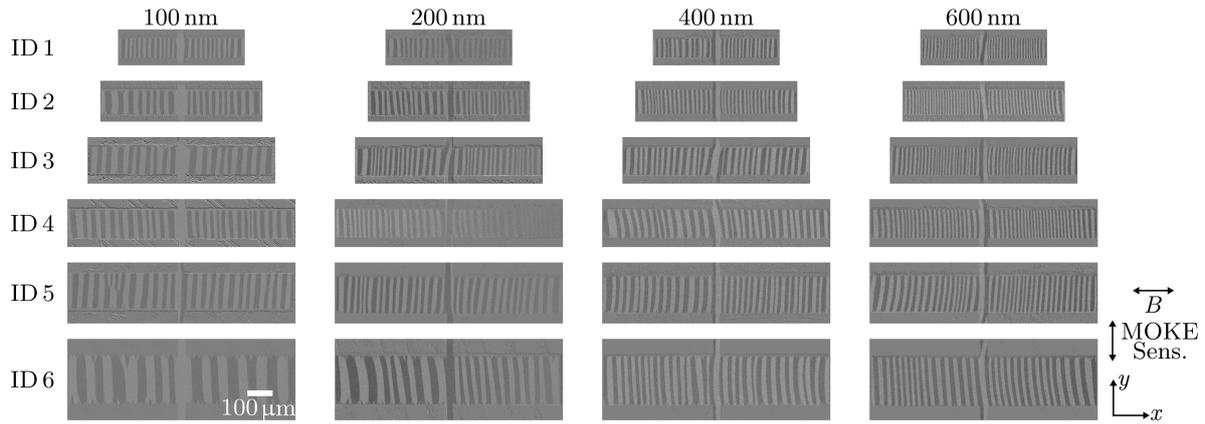

**Figure S1.** Magnetic domain images of the investigated sensors with FeCoSiB layer thicknesses of $t_\mathrm{m} =$ 100, 200, 400, and 600 nm across all geometries (ID 1 to ID 6) after demagnetizing the sensor along its long axis. The magnetooptical sensitivity is aligned perpendicular to the demagnetizing field.

To evaluate the contribution of shape anisotropy to the effective anisotropy of the sensors, the in-plane demagnetizing factor $D_\mathrm{x}$ along the length of the resonators is calculated for different magnetic layer thicknesses using an analytical model for thin rectangular plates.[1-2]

The results, shown in **Figure S3**, demonstrate that $D_x$ remains in the range of $10^{-9}$ to $10^{-7}$ at the center of the sensors and increases toward the edges. The resulting shape anisotropy energy densities are significantly smaller than the total effective anisotropy energy densities extracted from experiments. Therefore, the contribution of demagnetizing fields is negligible and cannot explain the observed reduction in magnetic susceptibility with increasing thickness.

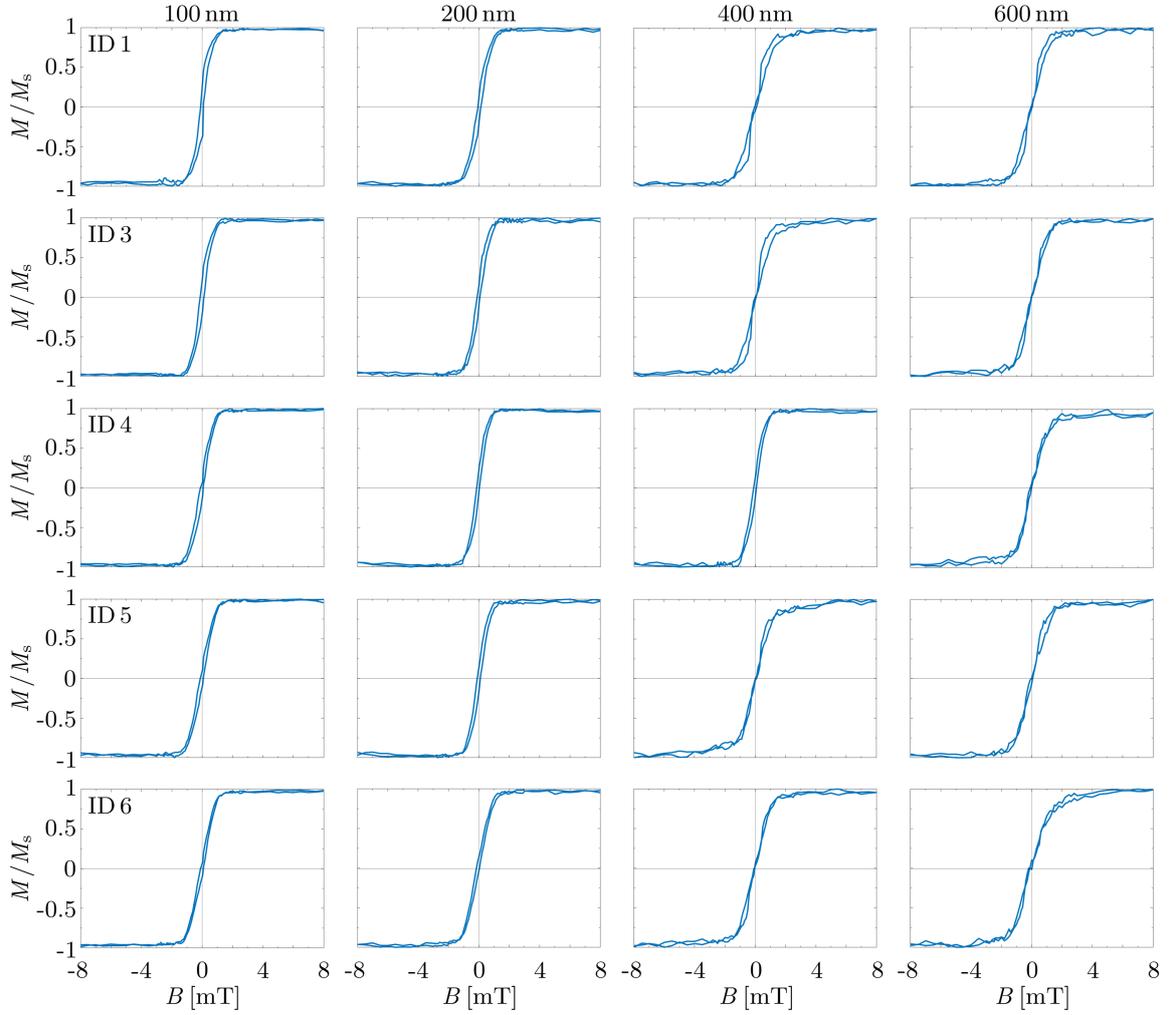

**Figure S2.** Magnetization curves of the investigated sensors with FeCoSiB layer thicknesses $t_m$ = 100, 200, 400, and 600 nm for geometries ID 1 and ID 3-6, measured along their long axes.

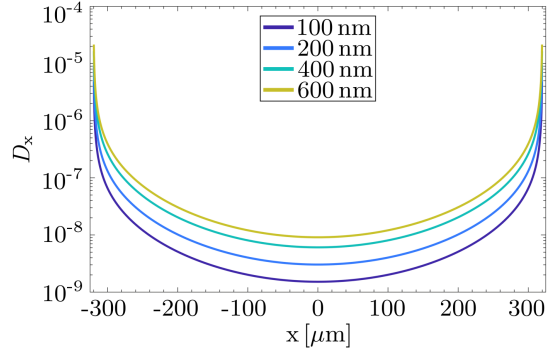

**Figure S3.** Calculated in-plane demagnetizing factor $D_x$ along the length of the resonators for different magnetic layer thicknesses ($t_m$ = 100, 200, 400, and 600 nm), based on an analytical model for thin rectangular plates. While $D_x$ is increasing at the edges, it remains in the range of $10^{-9}$ to $10^{-7}$ near the center of the sensors, indicating a negligible contribution of shape anisotropy to the overall effective anisotropy.

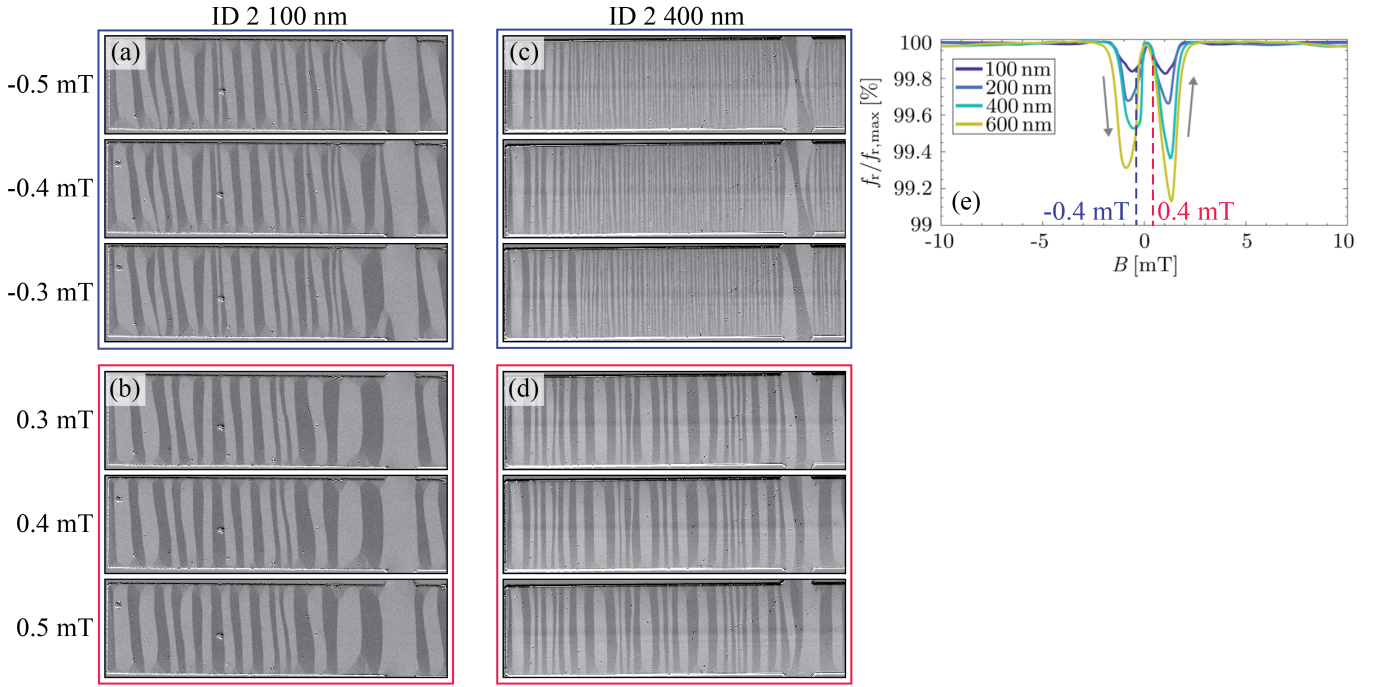

**Figure S4.** Selected magnetooptical images of the ID 2 samples with 100-nm-thick and 400-nm-thick magnetostrictive layer. Images (a) and (c) taken around -0.4 mT after saturation in the negative field illustrate the difference in the formation of the blocked narrow-domain state between the samples. Images (b) and (d) taken around 0.4 mT show the wide-domain state at the operation point. Similarity of the states (a) and (b) results in the symmetrical $f_r(B)$ curve (e). The prominent blocked state in (c) results in the asymmetrical $f_r(B)$.

**Electromechanical Properties**

**Figure S5** presents the admittance curves near the inflection points of sensor ID 2 at $B = 0.76$ mT for $\hat{u}_{ex} = 10$ mV and $\hat{u}_{ex} = 300$ mV for different FeCoSiB thicknesses, along with fitted mBvD models. At $\hat{u}_{ex} = 300$ mV, the admittance curves shift to lower frequencies compared to $\hat{u}_{ex} = 10$ mV for all thicknesses except $t_m = 100$ nm, where no shift is observed. The degree of the shift increases with increasing FeCoSiB thickness. Moreover, the curves exhibit greater asymmetry with increasing thickness, with significantly altered shapes and higher admittance magnitudes observed at $t_m = 600$ nm. While the linear mBvD model fits the measurements well at $\hat{u}_{ex} = 10$ mV, deviations between the model and measurements become apparent at $\hat{u}_{ex} = 300$ mV for thicker FeCoSiB layers, reflecting the onset of magnetoelastic nonlinearities.

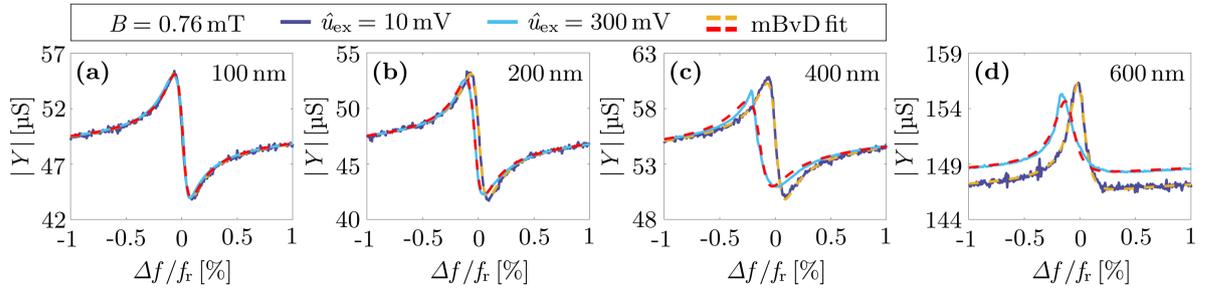

**Figure S5.** Measured admittance magnitude and simulations with a mBvD model at magnetic flux density $B = 0.76$ mT as a function of the normalized excitation frequency $\Delta f/f_r$ ($\Delta f \coloneqq f_{ex} - f_r$) at excitation voltage amplitudes $\hat{u}_{ex} = 10$ mV and $\hat{u}_{ex} = 300$ mV for FeCoSiB thicknesses of (a) 100 nm, (b) 200 nm, (c) 400 nm, and (d) 600 nm at RM3 of sensor ID 2

**Figure S6** complements the $f_r(B)$ curves, relative magnetic sensitivities $S_{m,r}$, relative electrical sensitivities $S_{el,r}$ and quality factors $Q$ discussed in the main text by by summarizing other electromechanical parameters as functions of normalized magnetic volume $Vol_{norm} \coloneqq Vol_{mag}/Vol_{mag,max}$, evaluated at the inflection point $B_{inf}$ of the inner slope on the positive side of the $f_r(B)$ curves at an excitation voltage amplitude of $\hat{u}_{ex} = 10$ mV. Figure S4a presents the total amplitude sensitivity $S_{am}$, which shows an overall increasing trend with increasing FeCoSiB thickness in most of the thickness and geometries, but deviations occur due to its dependence on both $S_{m,r}$ and $S_{el,r}$. Figure S4b illustrates the normalized frequency detuning $\Delta f_{r,norm} \coloneqq (f_{r,max} - f_{r,min})/f_{r,max}$, which generally increases with increasing magnetic volume. Figure S4c shows the magnetic bias flux density at the inflection points $B_{inf}$, which also tends to increase with increasing FeCoSiB thickness.

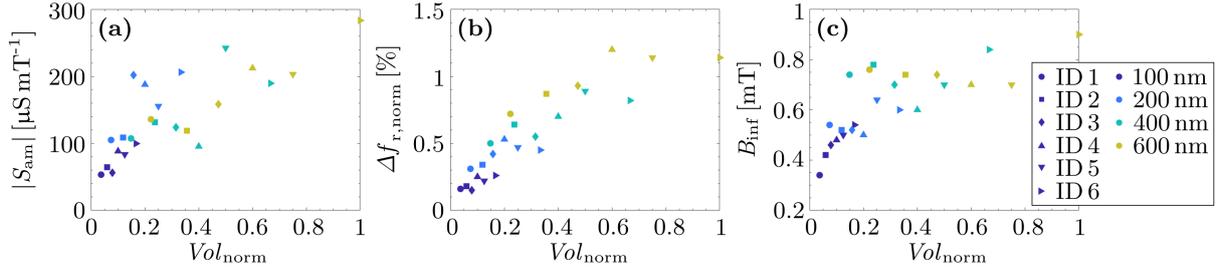

**Figure S6.** (a) Total amplitude sensitivity $S_{am}$, (b) normalized frequency detuning $\Delta f_{r,norm} \coloneqq (f_{r,max} - f_{r,min})/f_{r,max}$ and (c) magnetic flux density $B_{inf}$ corresponding to the inflection point of the inner slope on the positive side of the $f_r(B)$ curves as functions of normalized magnetic volume $Vol_{norm} \coloneqq Vol_{mag}/Vol_{mag,max}$ for all investigated sensors at an excitation voltage amplitude $\hat{u}_{ex} = 10$ mV.

**Signal and Noise Analysis**

**Figure S7** extends the comparison of measured and simulated performance of sensor ID2 to FeCoSiB thicknesses of 200 nm and 400 nm, complementing the 100 nm and 600 nm data presented in the main manuscript (Figure 5). Voltage sensitivity $S_V$, amplitude spectral density $E_{co}$, and limit of detection $LoD$ are plotted as functions of excitation voltage amplitude $\hat{u}_{ex}$. For all thicknesses, simulations predict a linear increase in $S_V$ and $E_{co}$ with $\hat{u}_{ex}$, assuming constant $S_{m,r}$ and $S_{el,r}$. However, the measurements begin to deviate from this linear trend at lower $\hat{u}_{ex}$ values for thicker films, indicating the earlier onset of nonlinearities. As observed, the $LoD$ initially decreases and then diverges from the model at higher $\hat{u}_{ex}$. The increasing discrepancy with thickness, as quantified by RMSE in the main text, confirms that nonlinear effects and magnetic noise contributions become more significant in thicker FeCoSiB films.

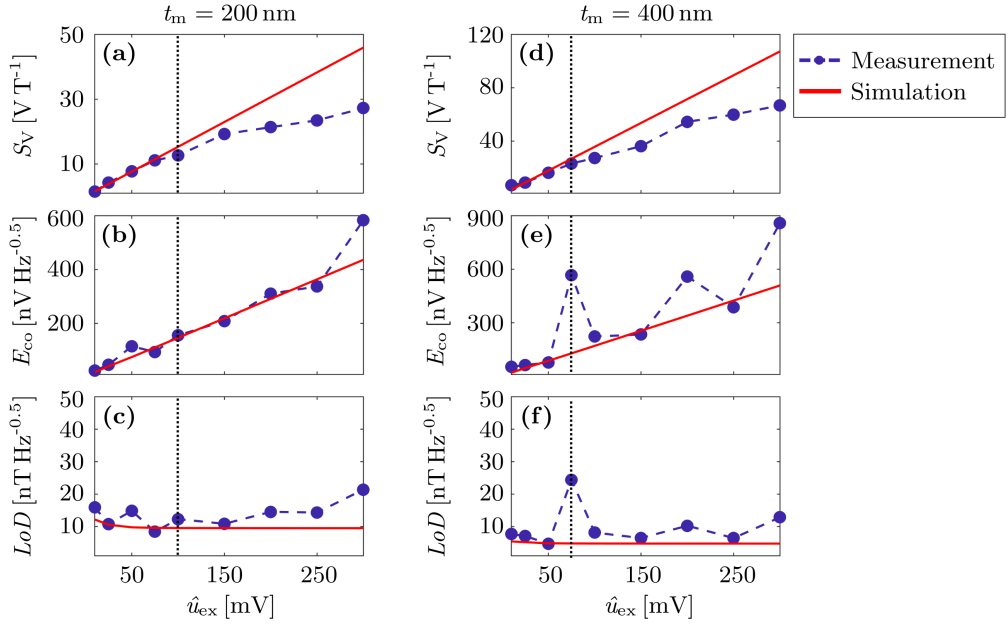

**Figure S7.** Comparison of measured and simulated voltage sensitivity $S_V$, amplitude spectral density $E_{co}$, and limit of detection $LoD$ of sensor ID 2 with FeCoSiB layer thicknesses of (a–c) 200 nm, and (d–f) 400 nm for a magnetic signal with an amplitude of $\hat{B}_{ac} = 1\,\mu\text{T}$ and a frequency of $f_{ac} = 10\,\text{Hz}$ as a function of excitation voltage amplitude $\hat{u}_{ex}$. Measurements and simulations are conducted at a magnetic flux density $B_{inf}$ corresponding to the inflection point of the inner slope on the positive side of the $f_r(B)$ curves. The black dotted vertical lines indicate the excitation voltage amplitude $\hat{u}_{ex,onset}$, corresponding to the onset of nonlinearities.

**References**

[1] A. Aharoni, *J. Appl. Phys.* **1998**, *83*, 3432.
[2] A. Aharoni, *Phys. status solidi* **2002**, *229*, 1413.